\documentclass[10pt,a4paper]{article}

\usepackage{graphicx}% Include figure files
\usepackage{dcolumn}% Align table columns on decimal point
\usepackage{bm}% bold math

\newcommand{\im}[1]{\textrm{Im}\left\{  #1\right\}}

\newcommand{\de}{\partial}
\newcommand{\sech}[1]{\textrm{sech}\left(  #1\right)}
\newcommand{\eq}[2]{\begin{equation} \label{#1} #2 \end{equation}}
\newcommand{\eps}{\epsilon}

\newcommand{\diag}{\textrm{diag}}
\newcommand{\etal}{{\em et al.}}

\newcommand{\cc}{\textrm{c.c.}}

\newcommand{\EE}{\mathbf{E}}
\newcommand{\BB}{\mathbf{B}}

\newcommand{\PP}{\mathbf{P}}

\title{\bf Gap solitons in spatiotemporal photonic crystals}

\author{Fabio Biancalana \\ {\em Department of Physics and Astronomy, Cardiff
University, Cardiff (UK)} \\ \\ Andreas Amann, Eoin P. O'Reilly \\
{\em Tyndall National Institute, Cork (Ireland)}}

\begin{document}

\maketitle

\begin{abstract}
We generalize the concept of nonlinear periodic structures to
systems that show arbitrary spacetime variations of the refractive
index. Nonlinear pulse propagation through these spatiotemporal
photonic crystals can be described, for shallow nonstationary
gratings, by coupled mode equations which are a generalization of
the traditional equations used for stationary photonic crystals.
Novel gap soliton solutions are found by solving a modified massive
Thirring model. They represent the missing link between the gap
solitons in static photonic crystals and resonance solitons found in
dynamic gratings.
\end{abstract}

The ability to manipulate spectrally and temporally optical pulses
has been a longstanding goal in modern science and technology, and
periodic media have the potential of engineering light propagation
to an unprecedented degree \cite{slusher}. Systems possessing a
periodic modulation of refractive index, in which photonic bandgaps
(PBGs) form at or near a multiple of the Bragg frequency (or
wavenumber), have been used in a wide range of applications,
including dispersion compensators and optical filters
\cite{slusher}. When Kerr nonlinearity is considered, new effects
come into play, such as optical bistability, pulse compression,
optical switching and soliton formation \cite{slusher}.

Stationary gap solitons (GSs) living in the {\em frequency} bandgap
of a 1D periodic medium  were first investigated in 1987-1989 in a
series of fundamental papers \cite{chenmills,christo,aceveswabnitz},
and found experimentally in 1996 \cite{eggleton}. Solitons living in
a {\em wavenumber} bandgap of the so-called dynamic gratings (i.e. a
traveling-wave periodic index change) were also investigated in
Refs. \cite{wabnitz,pitois,wavenumberbandgap} by using copropagating
beams, where the complementarity between the two kinds of bandgap
was evident. Since then, there has been an exponential increase in
the number of studies and practical applications of GSs. An
intriguing possibility is the storage of optical pulses in the form
of zero velocity GSs followed by release from the structure at a
controllable delay \cite{slusher}.

Much less attention, however, has been devoted to the physics of
{\em nonstationary} periodic media, such as dielectric structures
showing temporal variations of the refractive index
\cite{morgenthaler,biancalana}, which have the potential to
dramatically enhance the degree of spectral control over light
pulses by periodic media thanks to the new temporal degree of
freedom \cite{morgenthaler}. In Ref. \cite{biancalana} we derived
the transfer matrix $\mathcal{T}$ for plane waves scattered by the
sharp boundary associated with a medium with time-varying refractive
index, which must be distinguished from a moving interface in that
the medium itself is immobile \cite{biancalana,jackson}. The
knowledge of $\mathcal{T}$ for a single boundary allowed us to
construct a theory for more complicated nonstationary dielectric
objects. In particular in \cite{biancalana} we introduced the
important concept of {\em spatiotemporal photonic crystal} (STPC),
which is a grating that shows a well-defined periodicity of the
refractive index along a certain direction of the spacetime plane
$(z,ct)$ ($z$ is the longitudinal spatial coordinate, $t$ is time
and $c$ is the speed of light). This periodicity gives rise to PBGs
in a {\em mixed frequency-wavenumber} space, the mixing being
regulated by an angular parameter $\theta$, which we shall see it is
related to the apparent velocity of the layers in the spacetime
plane.

In this Letter we extend the linear theory formulated in Ref.
\cite{biancalana} to nonstationary gratings with Kerr nonlinearity,
demonstrating the existence of self-localized solutions in the mixed
frequency-wavenumber bandgaps of STPCs, thus showing that the
conventional concept of GS (see Refs.
\cite{chenmills,christo,aceveswabnitz,wabnitz,wavenumberbandgap})
must be extended to encompass general spacetime variations of the
refractive index.

%---------------------------------------------------------------------------------

Let us consider an electromagnetic wave, with its electric and
magnetic fields $\EE=(E(z,t),0,0)$ and $\BB=(0,B(z,t),0)$ linearly
polarized along the $\hat{x}$ and $\hat{y}$ directions respectively.
$E$ and $B$ depend on $z$ and $t$ only, because we assume conditions
of normal incidence, so that any change of the time-dependent
refractive index occurs along $\hat{z}$. The linear polarization
$\PP_{L}$ of the medium is given by
$\PP_{L}=\chi_{L}(z,t)\EE=[n(z,t)^{2}-1]\EE$, where $\chi_{L}$ is
the linear susceptibility of the (non-magnetic) medium and $n(z,t)$
is the linear refractive index, which is assumed for simplicity to
be real and frequency independent, and possessing for the moment an
arbitrary dependence on $z$ and $t$. Maxwell's equations for $E$ and
$B$ are $\de_{t}(\eps E)+c\de_{z}B+\de_{t}P_{NL}=0$,
$\de_{t}B+c\de_{z}E=0$, where $\eps(z,t)\equiv
1+\chi_{L}=n(z,t)^{2}$, and $P_{NL}$ is the Kerr nonlinear
polarization, $P_{NL}=\chi_{NL}E^{3}$, with $\chi_{NL}$ constant.
Here and in the following we use the Heaviside-Lorentz units system,
see Ref. \cite{jackson}.

\begin{figure}
\includegraphics{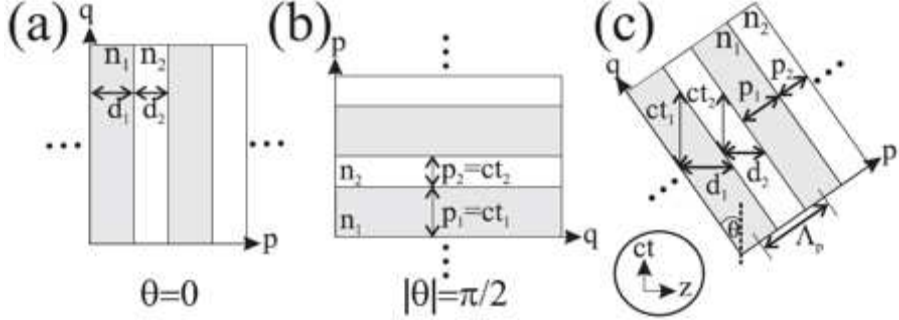}
\caption{\label{fig1} (a) Space-like, conventional static photonic
crystal ($\theta=0$), for which $p=z$ and $q=ct$. $n_{1,2}$ and
$d_{1,2}$ are respectively refractive indices and widths of the two
types of layers. (b) Time-like photonic crystal, $\theta=\pm\pi/2$,
for which the refractive index changes periodically in time only
($p=-ct$), and $q=z$. $ct_{1,2}$ are the durations of the layers.
(c) Intermediate case $0<|\theta|<\pi/2$. $\Lambda_{p}$ is the
period of the structure along the grating direction. Axis $ct$ and
$z$ are indicated in a circle.}
\end{figure}

By deriving the first of Maxwell's equations with respect to $t$,
and using the second one to eliminate $B$, we obtain the nonlinear
wave equation for a space- and time-varying refractive index:
\eq{maxwell2}{\eps\de_{t}^{2}E-c^{2}\de_{z}^{2}E+
(\de_{t}^{2}\eps)E+2(\de_{t}\eps)(\de_{t}E)+\de_{t}^{2}P_{NL}=0.} It
is now convenient to introduce two new variables, rotated by an
angle $\theta\in[-\pi/2,\pi/2]$ in the $(z,ct)$ plane:
$p=\cos(\theta)z-\sin(\theta)ct, q=\sin(\theta)z+\cos(\theta)ct$.
This spacetime rotation is analogous to the Lorentz transformations
in special relativity, with the essential conceptual difference that
in our case the associated dimensionless velocity $\beta\equiv
\tan(\theta)$ can assume values in the range $0<|\beta|<\infty$, and
it is not limited by $c$, see Ref. \cite{biancalana} and references
therein. $p$ will be chosen to correspond to the direction parallel
to the periodicity of the STPC, while $q$ will be orthogonal to $p$.
$\theta>0$ implies that the boundaries of the STPC are moving
towards light, while for $\theta<0$ the boundaries are moving away
from the incident pulse. A generalized plane wave propagating in the
$(p,q)$ space has the form
$\Psi=\Psi_{0}\exp(i\tilde{k}p-i\tilde{\omega}q)$, where $\Psi_{0}$
is a constant amplitude, and $\tilde{k}$ and $\tilde{\omega}$ are
the wavenumbers associated to the $p$ and $q$ directions
respectively. $\tilde{k}$ and $\tilde{\omega}$ are linked by the
rotation $\tilde{k}=\cos(\theta)k+\sin(\theta)\omega/c$ and
$\tilde{\omega}=\cos(\theta)\omega/c-\sin(\theta)k$, where $k$ and
$\omega/c$ are the wavenumbers associated to the original physical
plane $(z,ct)$. Figure \ref{fig1} shows the geometrical meaning of
axes $p$ and $q$ for three representative STPCs of fundamental
importance (see caption). It is evident from the above definitions
that the case $\theta\rightarrow 0$ corresponds to layers arranged
periodically along $z$ ($p\rightarrow z$), and the plane wave
delocalization direction lies along $ct$ ($q\rightarrow ct$). This
corresponds to the traditional time-independent photonic crystal,
see Fig. \ref{fig1}(a). Being the crystal invariant with respect to
translations along $ct$, we name this structure a {\em space-like}
STPC. More interesting is the second limiting case, when
$\theta\rightarrow\pm\pi/2$, shown in Fig. \ref{fig1}(b). From the
definitions of $p$ and $q$, we have $p\rightarrow \mp ct$ and
$q\rightarrow\pm z$, so that plane waves will be delocalized along
$z$, and all the variations of refractive index occur in time only.
In analogy with the previous nomenclature, we name this structure a
{\em time-like} STPC, an example of which is the dynamic grating
\cite{pitois,wabnitz}. The intermediate cases when
$0<|\theta|<\pi/2$, which are the main focus of this Letter, are
displayed schematically in Fig. \ref{fig1}(c).

%---------------------------------------------------------------------------------

In principle, integration of Eq.(\ref{maxwell2}) is all one needs to
completely solve the problem of nonlinear pulse propagation in any
kind of nonstationary dispersionless structure. However, one can
gain important analytical insight by considering a cosinusoidal
shallow grating described by the dielectric function
$\eps(p)=n_{0}^{2}+\mu[\exp(i\tilde{k}_{B}p)+\cc]$, where
$n_{0}^{2}$ is the square of the average linear refractive index,
and $\mu\ll n_{0}^{2}$. Here, $\tilde{k}_{B}$ represents the
equivalent of the Bragg wavenumber along $p$. The Bragg condition,
which is the phase-matching condition between the optical and the
grating wavenumbers, is given by
$\tilde{k}_{B}=\tilde{k}^{+}-\tilde{k}^{-}$. Due to the fact that
only two Fourier modes are present in the above expression for
$\eps(p)$, we can assume that only two optical modes strongly
contribute to the propagation dynamics, which leads us to the
expansion $E(p,q)=[F(p,q)e^{i\tilde{k}^{+}p-i\tilde{\omega}
q}+B(p,q)e^{i\tilde{k}^{-}p-i\tilde{\omega} q}+\cc]/2$, where $F$
and $B$ are envelopes of respectively the forward and backward
components of the electric field, $\tilde{k}^{\pm}$ are the
wavenumbers along $p$ for $F$ and $B$ respectively, and
$\tilde{\omega}$ is the wavenumber along $q$, which is common for
both components. Note that: (i) here the terms 'forward' and
'backward' are not in general associated to the spatial motion of
the modes (unless $\theta=0$), but rather to the more general motion
along the $p$-direction, and (ii) the $F$ and $B$ modes will
generally have different linear wavenumbers $|\tilde{k}^{\pm}|$
along $p$. The dispersion relation between $\tilde{k}^{\pm}$ and
$\tilde{\omega}$ is (see also \cite{biancalana})
$\tilde{k}^{\pm}=[\sin(\theta)\pm
n_{0}\cos(\theta)]\tilde{\omega}/[\cos(\theta)\mp
n_{0}\sin(\theta)]$, which is not valid for
$\theta=\pm\arctan(1/n_{0})$. For those angles, either $k^{+}$ or
$k^{-}$ diverge and the wavenumber along the $p$-direction is not
defined. Physically this is due to the fact that for
$\theta<-\arctan(1/n_{0})$ the layers boundaries of the STPC are
changing faster than the speed of light in the medium ($c/n_{0}$).

%---------------------------------------------------------------------------------

\begin{figure}
\includegraphics{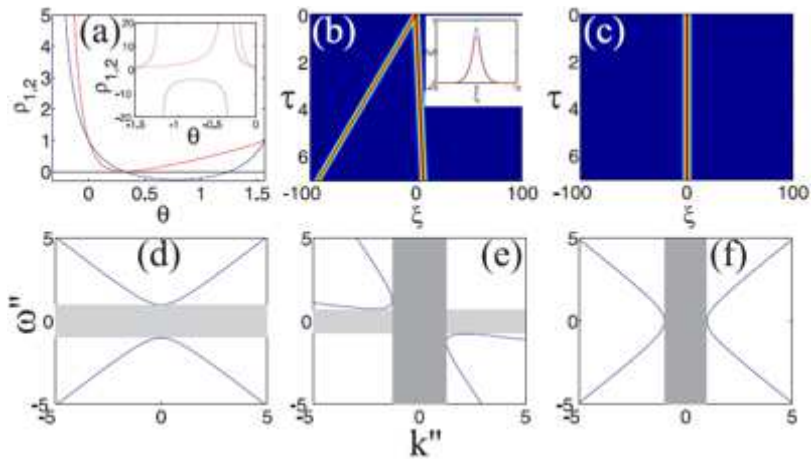}
\caption{\label{fig2} (Color online) (a) $\theta$-dependence of
coefficients $\rho_{1,2}$ for the parameter range
$\theta\in[-\arctan(1/n_{0}),\pi/2]$, and for $\theta\in[-\pi/2,0]$
(inset). (b) Contour plot of $I_{tot}=|f|^{2}+|b|^{2}$ for numerical
propagation of an initial STGP (input profile shown inset, blue line
is $|f|$ and red line is $|b|$) with parameters $v=0$, $\theta=1.3$
and $\delta=\pi/2$, without grating ($\kappa=0$). Propagation length
is $\tau=7$. (c) Same as (b) but with grating coupling. (d)
Bandstructure in space $(k'',\omega'')$ as calculated by using
(\ref{cme1}-\ref{cme2}), for $\theta=0$, (e) for $\theta=1.3$ and
(f) for $\theta=\pi/2$. In (c,d,e), light gray regions indicate the
$\omega''$-bandgap, and the dark gray regions the $k''$-bandgap. In
all cases $n_{0}=3$.}
\end{figure}

After substituting the expansion for $E(p,q)$ into
Eq.(\ref{maxwell2}), a slowly-varying amplitude approximation (SVEA)
in $p$ and $q$ is performed: $|\de_{p}^{2}\psi|\ll
|\tilde{k}^{\pm}\de_{p}\psi|\ll |(\tilde{k}^{\pm})^{2}
\psi|,|\de_{q}^{2}\psi|\ll |\tilde{\omega}\de_{q}\psi|\ll
|\tilde{\omega}^{2}\psi|$, where $\psi$ is either $F$ or $B$, and
similar relations are valid for the terms containing the mixed
derivative $\de_{p}\de_{q}$. The following two {\em spatiotemporal
coupled mode equations} (STCMEs) are obtained:
\begin{eqnarray}
\label{cme11}
i\de_{p}F+\frac{n_{0}\cos(\theta)+\sin(\theta)}{\cos(\theta)-n_{0}\sin(\theta)}i\de_{q}F+
\frac{\kappa}{[\cos(\theta)-n_{0}\sin(\theta)]^{2}}B+
\frac{\Gamma}{[\cos(\theta)-n_{0}\sin(\theta)]^{2}}(|F|^{2}+2|B|^{2})F=0,
\\
\label{cme22}
-i\de_{p}B+\frac{n_{0}\cos(\theta)-\sin(\theta)}{\cos(\theta)+n_{0}\sin(\theta)}i\de_{q}B+
\frac{\kappa}{[\cos(\theta)+n_{0}\sin(\theta)]^{2}}F+
\frac{\Gamma}{[\cos(\theta)+n_{0}\sin(\theta)]^{2}}(2|F|^{2}+|B|^{2})B=0,
\end{eqnarray}
where $\Gamma\equiv(3\tilde{\omega}\chi_{NL})/(8n_{0})$ is the
nonlinear coefficient, and $\kappa\equiv \tilde{\omega}\mu/(2n_{0})$
is the grating coupling constant. Again, the equations have a
singular character for $\theta=\pm\arctan(1/n_{0})$.
Eqs.(\ref{cme11}-\ref{cme22}) represent the first central result of
this Letter. Let us now perform the following scaling, which is
well-defined for $\theta\neq\arctan(1/n_{0})$ and
$\theta\neq-\arctan(n_{0})$: $\tau=q/q_{0}$, $\xi=p/p_{0}$,
$f=F/A_{0}$, $b=B/A_{0}$, with
$p_{0}\equiv\kappa^{-1}[\cos(\theta)-n_{0}\sin(\theta)]^{2}$,
$q_{0}\equiv[n_{0}\cos(\theta)+\sin(\theta)]p_{0}/[\cos(\theta)-n_{0}\sin(\theta)]$,
$F_{0}\equiv(\kappa/\Gamma)^{1/2}$. With this, equations
(\ref{cme11}-\ref{cme22}) are reduced to the following two
dimensionless equations:
\begin{eqnarray} \label{cme1} i\left(
\de_{\xi}+\de_{\tau} \right)f+b+\left( 2|b|^{2}+|f|^{2} \right)f=0,
\\ \label{cme2} i\left( -\de_{\xi}+\rho_{1}\de_{\tau}
\right)b+\rho_{2}f+\rho_{2}\left( 2|f|^{2}+|b|^{2} \right)b=0,
\end{eqnarray}
where
$\rho_{1}(\theta)\equiv[n_{0}\cos(\theta)-\sin(\theta)][\cos(\theta)-n_{0}\sin(\theta)]/
\{[\cos(\theta)+n_{0}\sin(\theta)][n_{0}\cos(\theta)+\sin(\theta)]\}$
and $\rho_{2}(\theta)
\equiv\{[\cos(\theta)-n_{0}\sin(\theta)]/[\cos(\theta)+n_{0}\sin(\theta)]\}^{2}$.
In Figure \ref{fig2}(a) coefficients $\rho_{1,2}$ are shown as a
function of $\theta$ for an average refractive index $n_{0}=3$. Note
that although $\rho_{2}$ is always positive in the range
$\theta\in[-\pi/2,\pi/2]$, $\rho_{1}$ becomes negative for
$\arctan(1/n_{0})<|\theta|<\arctan(n_{0})$, and shows two
divergences for negative angles at $\theta=-\arctan(n_{0})$ and at
$\theta=-\arctan(1/n_{0})$. Also $\rho_{1}=\rho_{2}=1$ for the
limiting cases $\theta=0$ and $\theta=\pi/2$, but in general these
parameters can strongly differ from unity.

%---------------------------------------------------------------------------------

Let us now discuss the most important linear property of
Eqs.(\ref{cme1}-\ref{cme2}), namely the PBG in the $\theta$-rotated
frequency-wavenumber space. Substituting
$\{f,b\}=\Psi_{f,b}\exp(ik'\xi-i\omega'\tau)$ into
Eqs.(\ref{cme1}-\ref{cme2}), and neglecting the nonlinear terms, we
readily obtain
$\omega'_{1,2}(\theta)=\{(\rho_{1}-1)k'\pm[(1+\rho_{1})^{2}k'^{2}+4\rho_{1}\rho_{2}]^{1/2}\}/(2\rho_{1})$,
after which we perform an inverse rotation back to the original
dimensionless frequency-wavenumber space, i.e.
$k''=\cos(\theta)k'-\sin(\theta)\omega'$,
$\omega''=\sin(\theta)k'+\cos(\theta)\omega'$. In Figure
\ref{fig2}(d,e,f) the bandstructure $\omega''_{1,2}(k'')$ for three
different cases ($\theta=0$, $\theta=1.3$ and $\theta=\pi/2$) is
plotted, explicitly showing the passage from the frequency bandgap
[$\theta=0$, Fig. \ref{fig2}(d)] to the wavenumber bandgap
[$\theta=\pi/2$, Fig. \ref{fig2}(f)], passing through a region in
which the two kinds of bandgap coexist [$\theta=1.3$, Fig.
\ref{fig2}(e)].

%---------------------------------------------------------------------------------

We now proceed to analyze the symmetries and the GS solutions of
Eqs.(\ref{cme1}-\ref{cme2}). One can derive
Eqs.(\ref{cme1}-\ref{cme2}) from the following Hamiltonian density:
$\mathcal{H}=(bf^{*}+fb^{*})+2|f|^{2}|b|^{2}+(|f|^{4}+|b|^{4})/2-\mathcal{M}_{f}
+\mathcal{M}_{b}/\rho_{2}$, where the star indicates complex
conjugation, and where $\mathcal{M}_{\zeta}\equiv
i(\zeta\de_{\xi}\zeta^{*}-\cc)/2=\im{\zeta^{*}\de_{\xi}\zeta}$ is
the momentum density of the generic field $\zeta=\{ f,b \}$. The
dynamical equations are written as
$i\hat{J}\hat{M}\de_{\tau}\mathbf{g}+\delta\mathcal{H}/\delta\mathbf{g}^{\dagger}=0$,
where $\hat{M}\equiv\diag(1,\rho_{1}/\rho_{2})$,
$\hat{J}\equiv\diag(1,-1)$ is the symplectic matrix,
$\mathbf{g}\equiv [f,b]$ is the field vector, and dagger indicates
hermitian conjugation. Moreover, the variational derivative is given
by
$\delta/(\delta\zeta)\equiv\de/(\de\zeta)-\de_{\xi}[\de/\de(\de_{\xi}\zeta)]$,
see also Ref. \cite{morrison}. We anticipate that $\xi$ will
correspond to the localization coordinate of the soliton solutions,
and $\tau$ to the evolution coordinate. With this in mind, one can
find the total Hamiltonian by integrating $\mathcal{H}$ over $\xi$,
$H=\int_{-\infty}^{+\infty}\mathcal{H}d\xi\equiv[\mathcal{H}]_{-\infty}^{+\infty}$.
$H$ is an integral of motion, i.e. $\de_{\tau}H=0$. $\mathcal{H}$
does not depend on the variable $\xi$ explicitly, leading to the
conservation of total momentum:
$M_{tot}=[\mathcal{M}_{f}+(\rho_{1}/\rho_{2})\mathcal{M}_{b}]_{-\infty}^{+\infty}$,
i.e. $\de_{\tau}M_{tot}=0$. $\mathcal{H}$ is also invariant with
respect to the 'gauge transformation' $f\rightarrow f\exp(i\phi)$,
$b\rightarrow b\exp(i\phi)$, leading to the conservation of the
quantity
$P=[|f|^{2}+(\rho_{1}/\rho_{2})|b|^{2}]_{-\infty}^{+\infty}$, and
$\de_{\tau}P=0$. The number of integrals of motion of the dynamical
system determined by (\ref{cme1}-\ref{cme2}) (with the exclusion of
$H$) is closely related to the number of internal parameters of the
corresponding soliton families \cite{parameters}. Therefore the
family of localized solutions living inside the
$\tilde{\omega}$-bandgap of a STPC are represented by two internal
parameters. This is well-known for solitons living in the frequency
bandgap of a static photonic crystal ($\theta=0$)
\cite{christo,aceveswabnitz}, and for GSs living in the wavenumber
bandgap ($\theta=\pm\pi/2$) \cite{wabnitz,wavenumberbandgap}, but
our analysis extends this result for arbitrary values of $\theta$.

%---------------------------------------------------------------------------------

In order to find analytical localized solutions of
Eqs.(\ref{cme1}-\ref{cme2}), let us now consider the following
different set of coupled equations for two new fields $\psi_{f}$ and
$\psi_{b}$:
\begin{eqnarray}
\label{mmtm1}i\left( \de_{\xi}+\de_{\tau}
\right)\psi_{f}+\psi_{b}+|\psi_{b}|^{2}\psi_{f}=0, \\
\label{mmtm2}i\left( -\de_{\xi}+\rho_{1}\de_{\tau}
\right)\psi_{b}+\rho_{2}\psi_{f}+\rho_{2}|\psi_{f}|^{2}\psi_{b}=0.
\end{eqnarray}
In analogy with the extensively studied Massive Thirring Model (MTM)
\cite{massivethirringmodel,aceveswabnitz}, we name
Eqs.(\ref{cme1}-\ref{cme2}) the {\em modified Massive Thirring
Model} (mMTM). The MTM is a particular case of the mMTM, with
$\rho_{1}=\rho_{2}=1$, and it is known to be integrable
\cite{massivethirringmodel}. The mMTM solitons will automatically
provide analytical soliton solutions to the original equations
Eqs.(\ref{cme1}-\ref{cme2}). Let us operate the Galileian shift
$\tau'=\tau-V\xi$, $\xi'=\xi$, and the rescalings
$\bar{\tau}=\tau'/\tau_{0}$ and $\bar{\xi}=\xi'$. By choosing
$\tau_{0}=1+V$ and $V=(1-\rho_{1})/2$ we can therefore scale
$\rho_{1}$ away from Eqs.(\ref{mmtm1}-\ref{mmtm2}). It is now
possible to find the following analytical soliton solution:
\begin{eqnarray} \label{sol4}
\psi_{f}(\bar{\xi},\bar{\tau})=(1/\Delta)\psi_{0},\qquad\psi_{b}(\bar{\xi},\bar{\tau})=-\Delta\psi_{0},
\end{eqnarray} where
$\psi_{0}=\sin(\delta)\sech{\Theta-i\delta/2}\exp(i\sigma)$,
$\gamma=\left[\rho_{2}/(1-v^{2})\right]^{1/2}$, $\Delta=
[\rho_{2}(1-v)/(1+v)]^{1/4}$,
$\Theta=\gamma\sin(\delta)(\bar{\xi}-v\bar{\tau})=\gamma\sin(\delta)\{
[1+vV/(1-V)]\xi-V\tau/(1-V)\}$ and
$\sigma=\gamma\cos(\delta)(v\bar{\xi}-\bar{\tau})=\gamma\cos(\delta)\{[v+V/(1-V)]\xi-\tau/(1-V)\}$.
$0<\delta<\pi$ is a parameter (also called the {\em soliton charge})
which measures the detuning from the bandgap center
($\delta=\pi/2$), and $-1\le v\le +1$ is the soliton relative
velocity.

%---------------------------------------------------------------------------------

\begin{figure}
\includegraphics{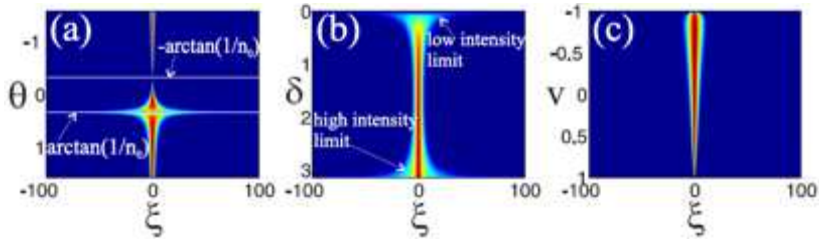}
\caption{\label{fig3} (Color online) Contour plots of the total
intensity profile $I_{tot}=|f|^{2}+|b|^{2}$ for the STGS: (a) as a
function of $\theta$, for $v=0$ and $\delta=\pi/2$ [note the
divergence located at $\theta=\arctan(1/n_{0})$, and the vanishing
intensity in correspondence of $\theta=-\arctan(1/n_{0})$], (b) as a
function of $\delta$, for $\theta=1.3$ and $v=0$, and (c) as a
function of $v$, for $\theta=1.3$ and $\delta=\pi/2$. In all cases
$n_{0}=3$.}
\end{figure}

We now attempt to express the general localized solutions of
Eqs.(\ref{cme1}-\ref{cme2}) in terms of mMTM solitons,
Eqs.(\ref{sol4}), by using the ansatz:
$\{f,b\}=\alpha\psi_{f,b}(\xi,\tau)\exp[i\eta(\Theta(\xi,\tau))]$.
Substituting into Eqs.(\ref{cme1}-\ref{cme2}) and using
(\ref{sol4}), we obtain two equations for
$\eta^{\prime}\equiv\de_{\Theta}\eta$. The consistency condition
between them determines the value of $\alpha$:
$\alpha=\{[2(1-v^{2})\rho_{2}]/[(1+v)^{2}+4(1-v^{2})\rho_{2}+(1-v)^{2}\rho_{2}^{2}]\}^{1/2}$,
which in turn is used to solve the ODE for $\eta(\Theta)$, obtaining
the solution: $e^{i\eta(\Theta)}=\{
[1+e^{i\delta+2\Theta}]/[e^{i\delta}+e^{2\Theta}]\}
^{[(1+v)^{2}-(1-v)^{2}\rho_{2}^{2}]/[(1+v)^{2}+4(1-v^{2})\rho_{2}+
(1-v)^{2}\rho_{2}^{2}]}$. This completes the information necessary
to find the two-parameter family of localized solutions, i.e. the
{\em spatiotemporal gap solitons} (STGSs), for the STCMEs given by
Eqs.(\ref{cme1}-\ref{cme2}), which represents the second central
result of this Letter. The intensity ratio $r$ between $f$ and $b$
is given by $r\equiv|f|^{2}/|b|^{2}=(1+v)/[(1-v)\rho_{2}]$, so that
$f$ and $b$ for the zero-velocity solitons ($v=0$) do not have in
general equal amplitudes [$r(v=0)=1/\rho_{2}$]. Figure \ref{fig3}
shows contour plots of the soliton total intensity
$I_{tot}=|f|^{2}+|b|^{2}$ when changing $\theta$ [Fig.
\ref{fig3}(a)], $\delta$ [Fig. \ref{fig3}(b)] and finally $v$ [Fig.
\ref{fig3}(c)].

%---------------------------------------------------------------------------------

Our analytical solution is also confirmed by direct numerical
integration of Eqs.(\ref{cme1}-\ref{cme2}), performed using a
split-step Fourier method with a $4$th order Runge-Kutta algorithm.
Figure \ref{fig2}(b,c) shows the propagation of a STGS with
parameters $v=0$, $\theta=1.3$ and $\delta=\pi/2$, which lives in
the center of the mixed bandgap displayed  in Fig. \ref{fig2}(e),
for a propagation of $\tau=7$. Fig. \ref{fig2}(b) shows that when
the grating is absent ($\kappa=0$) the two components separate and
do not interact, while Fig. \ref{fig2}(c) shows the undisturbed
soliton propagation at zero relative velocity in presence of the
spatiotemporal grating. Surprisingly, it is seen from the numerical
simulations that quasi-adiabatic variations of $\theta$ during
propagation, which thus change dynamically the background
spatiotemporal grating, do not destroy the STGS, due to prompt pulse
reshaping. This {\em structural stability} makes STGSs very
attractive for storing, slowing down, converting and releasing
optical energy in a controlled way, which may have profound
implications for optical communications and quantum information
processing \cite{lukin}.

%---------------------------------------------------------------------------------

In conclusion, in this Letter we have derived a set of CMEs that
allow to describe nonlinear pulse propagation in a shallow grating
with space-time variations of the refractive index. This structure
generally possesses a bandgap in a rotated frequency-wavenumber
space, where new GSs have been found analytically by solving an
associated mMTM. Our formulation considerably generalizes the
current theoretical understanding of periodic media to
time-dependent refractive index. Future works will include the
bifurcation and stability analysis of STGSs, and the natural
extension of the theory to dispersive media.

%---------------------------------------------------------------------------------

%\acknowledgements

We acknowledge financial support from the UK Engineering and
Physical Sciences Research Council (EPSRC), Science Foundation
Ireland (SFI) and the Irish Research Council for Science,
Engineering and Technology (IRCSET).


\begin{thebibliography}{99}
%\begin{references}

\bibitem{slusher} R. E. Slusher, B. J. Eggleton, {\em Nonlinear photonic crystals} (Springer, Berlin, 2003).

\bibitem{chenmills} W. Chen and D. L. Mills, Phys. Rev. Lett. {\bf
58}, 160 (1987).

\bibitem{christo} D. N. Christodoulides and R. I. Joseph, Phys. Rev.
Lett. {\bf 62}, 1746 (1989).

\bibitem{aceveswabnitz} A. B. Aceves and S. Wabnitz, Phys. Lett. A
{\bf 141}, 37 (1989).

\bibitem{eggleton} B. J. Eggleton \etal, Phys. Rev. Lett. {\bf 76}, 1627
(1996).

\bibitem{morgenthaler} F. R. Morgenthaler, IRE Trans. Microwave Theory Tech. {\bf MTT-6},
167 (1958).

\bibitem{biancalana} F. Biancalana \etal, Phys. Rev. E {\bf 75},
046607 (2007).

\bibitem{jackson} J. D. Jackson, {\em Classical Electrodynamics} (Wiley and Sons, New York, 1975).

\bibitem{wabnitz} S. Wabnitz, Opt. Lett. {\bf 14}, 1071 (1989).

\bibitem{wavenumberbandgap} G. Van Simaeys \etal, Phys. Rev. Lett. {\bf 92}, 223902 (2004).

\bibitem{pitois} S. Pitois, M. Haelterman and G. Millot, J. Opt. Soc. Am. B {\bf 19}, 782 (2002).

\bibitem{morrison} P. J. Morrison, Rev. Mod. Phys. {\bf 70}, 467
(1998).

\bibitem{parameters} K. A. Gorshkov and L. A. Ostrovsky, Physica D
{\bf 3}, 428 (1981).

\bibitem{massivethirringmodel} W. E. Thirring, Ann. Phys. (NY) {\bf
3}, 91 (1958); E. A. Kuznetsov and A. V. Mikhailov, Teor. Mat. Fiz.
{\bf 30}, 193 (1970).

\bibitem{lukin} M. D. Lukin and A. Imamoglu, Nature {\bf 413}, 273
(2001); L. M. Duan \etal, Nature {\bf 414}, 413 (2001).


\end{thebibliography}
\end{document}